\documentclass[useAMS,usenatbib]{mn2e}
\usepackage{graphicx,natbib}

\begin{document}
\newif\ifAMStwofonts
\AMStwofontstrue
%%%%%%%%%%%%%%%%%%%%%%%%%%%%%%%%%%%%%%%%%%%%%%%%

\newcommand{\kms}{~km~s$^{-1}$}
\newcommand{\vrot}{$v_{\rm rot}$}
\newcommand{\mdot}{$\rm{M}_{\odot}$}
\newcommand{\chisq}{$\chi^{2}$}
\newcommand{\vsini}{$v\sin i$}
\newcommand{\dg}{^{\circ}}
\newcommand{\ha}{{\mbox{H$\rm \alpha$}}}
\newcommand{\hb}{{\mbox{H$\rm \beta$}}}

\def\aap{Astron.~Astrophys.~}
\def\aaas{Astron.~Astrophys.~Suppl.~}
\def\mnras{Mon.~Not.~R.~Astron.~Soc.~}
\def\aas{Astron.~Astrophys.~}
\def\aas{Astron.~Astrophys.~Suppl.~}
\def\aj{Astron.~J.~}
\def\apj{Astrophys.~J.~}
\def\apjl{Astrophys.~J.~Lett.~}
\def\apjs{Astrophys.~J.~Suppl.~}
\def\baas{Bull.~Am.~Astron.~Soc.~}
\def\pasp{Publ.~Astron.~Soc.~Pac.~}
\def\pasj{Publ.~Astron.~Soc.~Jap.~}
\def\araa{Ann.~Rev.~Astron.~Astrophys.~}
\def\an{Astron.~Nachr.~}
\def\apss{Astrophys.~Sp.~Sci.~}
\def\aaps{Astron.~Astrophys.~Suppl.~}

\title{Doppler Images and Chromospheric Variability of TWA~17}
\author[M.~B.~Skelly et al]{M.~B.~Skelly,$^{1,4}$\thanks{E-mail: mskelly@ast.obs-mip.fr}  
Y.~C.~Unruh,$^{1}$ J. R. Barnes,$^{2}$ W. A. Lawson,$^{3}$  J.-F. Donati $^{4}$ 
\newauthor
A. Collier Cameron,$^{5}$ \\
$^{1}$ Astrophysics Group, Imperial College of Science, Technology and Medicine, London SW7 2AZ \\
$^{2}$ Centre for Astrophysics Research, University of Hertfordshire, College Lane, Hatfield, Herts AL10 9AB  \\
$^{3}$ School of Physical, Environmental and Mathematical Sciences, University of New South Wales, \\
Australian Defence Force Academy, Canberra, ACT 2600, Australia \\ 
$^{4}$ Observatoire Midi-Pyr\'{e}n\'{e}es, LATT, 14 avenue Edouard Belin, 31400 Toulouse, France \\
$^{5}$ School of Physics and Astronomy, University of St Andrews, Fife KY16 9SS \\  }
\date{16 July 2009}

\pagerange{\pageref{firstpage}--\pageref{lastpage}}
\pubyear{2009}

\newtheorem{theorem}{Theorum}[section]

\label{firstpage}
\maketitle

\begin{abstract}
We present Doppler imaging and a Balmer line analysis of the weak-line T Tauri star TWA~17. 
Spectra were taken in 2006 with the UCL Echelle Spectrograph on the Anglo-Australian 
Telescope. Using least-squares deconvolution to improve the effective signal-to-noise ratio we 
produced a Doppler map of the surface spot distribution. This shows similar features to 
maps of other rapidly rotating T 
Tauri stars, i.e. a polar spot with more spots extending out of it down to the equator. 

In addition to the photospheric variability, the chromospheric variability was studied using the Balmer emission. The mean H$\rm \alpha$ profile has a narrow component 
consistent with rotational broadening and a broad component extending out to 
$\pm 220$~\kms.
The variability in H$\rm \alpha$ suggests that the chromosphere has at least one slingshot prominence $3R_{*}$ above the surface.

% association

\end{abstract}


\section{Introduction}\label{intro}

The intrinsic variability and faintness of late-type pre-main sequence (PMS) stars has made it difficult to gain a complete picture of their development in the crucial first few tens of million years of their lives. In particular, this is because their variability often means that only long--term monitoring can give a useful picture of their behaviour. The tracks that young stars follow onto the main sequence are therefore not well defined observationally, and different theoretical PMS tracks can predict widely differing parameters. Comprehensive observations of stars at each stage of their evolution to the main-sequence are needed. The discovery in the last ten years of nearby co-moving groups of young stars has provided several populations of stars with similar ages and distances \citep{zuck04}  from which targets suitable for extensive study can be identified.

These groups include populations of T Tauri stars (TTS). Stars of spectral types F -- M are considered to be T Tauri stars after they become optically visible. They are still contracting, and stable hydrogen burning has not started. The radiated energy of the star at this time is due to its gravitational contraction. After the T Tauri phase has ended it may still be tens of millions of years before the star reaches the zero--age main sequence (ZAMS). During this time the star is in the post-T Tauri stage. 

A key feature of young stars is their magnetic activity which manifests itself, e.g., in the presence of large spots covering a substantial proportion of the photosphere. Starspots are formed when magnetic flux tubes emerge through the surface of the star, reducing convection to these points and causing a reduction in temperature and the formation of cool (and therefore dark) spots. In non-accreting T Tauri stars the presence of cool spots is the dominant cause of the variability and gives rise to lightcurves with clearly defined periods (equal to the rotational period) and amplitudes of up to a few tenths of a magnitude in the $V$ band. Doppler imaging \citep{vogt83,rice96,strass02,piskunov08} can be used to produce maps of the spots on rapidly rotating stars by taking advantage of the distortions the spots cause to rotationally broadened absorption lines. 

Prominences -- material supported by the magnetic field lines extending from the surface, are also a magnetic phenomenon. In young stars `slingshot' prominences have been observed, e.g. in Speedy Mic \citep{duns06,wolter08}, AB Dor \citep{cc89,abdor99} and TWA 6 \citep{skelly1}. These are clouds of material supported by the magnetic field at distances often beyond the corotation radius.  
The Balmer lines, in particular H$\rm \alpha$, can be used to study the chromosphere, and slingshot prominences show up as increased emission outside the projected rotational velocity and as rapidly moving absorption transients inside the projected rotational velocity. 

\subsection{TWA~17}

The TW Hydrae Association (TWA) is a nearby co-moving group with an estimated age of $10^{7} \rm ~yr$. Several of its members have Hipparcos distances around 50 pc from Earth. In \cite{lawson05} two subgroups were identified within the TWA;  the members numbered 1--13 lie at an average distance of $\sim 55 \rm pc$, and those numbered 14--19 probably belong to the Lower Centaurus--Crux (LCC) subgroup of the Ophiuchus--Scorpius--Centaurus OB-star association. A similar conclusion was reached by \cite{mamajek05} using the moving cluster method to estimate distances to individual TWA stars. 

In \cite{skelly1} (hereafter Paper~I) we discussed the weak-line T Tauri star TWA~6. Doppler images were presented and used to demonstrate that there is no differential rotation in this star. Using the Balmer line emission variability we showed evidence for the presence of large-scale prominences and active regions that were cospatial with the spots on the photosphere. 
In this paper we analyse another reported member of the same association - TWA~17. TWA~17 has a spectral type of K5, a period of 0.69 d \citep{lawson05}, and a \vsini\ of 45\kms\ \citep{webb99}. 

The observations and data analysis  of TWA~17 are set out in Sections \ref{obs} and \ref{analysis}. The Doppler images are presented in Section \ref{dopim}, followed by a discussion of the Balmer line emission in Section \ref{balmer}. In Section \ref{comp} we present arguments that TWA~17 in fact lies in the LCC subgroup.  

\section{Observations and Data Reduction}\label{obs}

The observations, carried out 2006 February 11--20, are described in detail in Paper~I. A brief summary is given here. 

Spectra were taken of TWA~6 and TWA~17 using the University College London Echelle Spectrometer (UCLES) at the Anglo-Australian Telescope (AAT). UCLES was centred at $5500\rm{\AA}$ giving a wavelength coverage of $4400 - 7200\rm{\AA}$. A slit width of 1.0\arcsec~was used, giving a spectral resolution of 45000. As well as the target stars, we observed template stars of a variety of spectral types, B stars to obtain the telluric/sky spectrum, flats, darks, biases and ThAr arc frames. Observations were made 2006 February 11--13 and 17--20, but the second set of nights was concentrated on TWA~6 in order to measure differential rotation in that star. The full observing log is given in Tables 1 and 2 of Paper I. 
During the first set of nights 27 usable spectra of TWA~17 were obtained, and two more were taken during the second set of nights. 

\section{Least-squares Deconvolution}\label{analysis}

All data were reduced using the {\sc starlink} routine {\sc echomop} \citep[more details in][and Paper I]{echomop1997}. To improve the effective signal-to-noise ratio (SNR) of the data we used the technique of least-squares deconvolution \citep[LSD, described in][]{jf97}. This technique treats a spectrum as a convolution of a line profile and a line-list (effectively a series of delta--functions). We produced a high SNR line profile from each spectrum by deconvolving the line-list, obtained from the Vienna Atomic Line Database \citep[VALD,][]{vald}, using the program {\sc spdecon} \citep[described in][]{barnes98}. 

2500 absorption lines were used in the deconvolution. The mean SNR of the input spectra was 18, and of the LSD profiles was 630, giving a gain of 35. The continuum normalisation was successful in most cases. This was evident from the fact that in each profile the continuum regions at either side of the deconvolved profiles were at the same level. The exceptions were the two profiles produced from the observations that took place during the second set of nights, indicating a small change in the instrumental set-up in the mean-time. This was rectified by fixing a slope to the lengths of continuum on either side of the profile and subtracting.  

\begin{figure}
\includegraphics[width=0.48\textwidth]{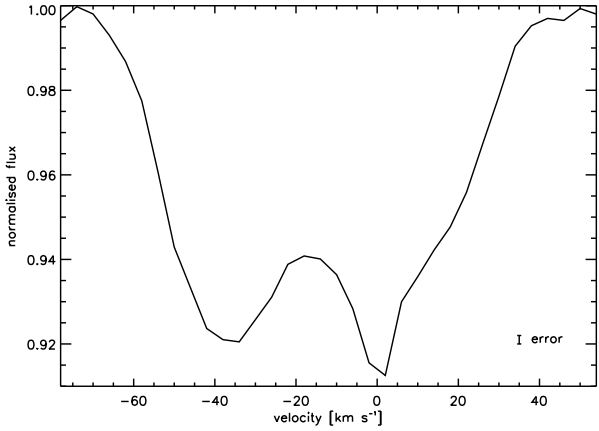}
\caption{Line profile at phase 0.736. The data points are at intervals of 4\kms\ and the magnitude of the errors is indicated by the small bar on the bottom right. The large feature at 0\kms\ due to the Moon is obvious.}
\label{exprof}
\end{figure}

In almost all of our spectra there was a strong absorption feature close to 0\kms\ (see Fig. \ref{exprof}). As discussed in Paper~I, this is due to the presence of scattered moonlight contaminating the spectra. As TWA~17 has a lower \vsini\ than TWA~6 the `Moon' profile covers a larger portion of the profiles. It is difficult to disentangle the size of this absorption feature from the changes to the profile due to spots so we used the following strategy, described in \cite{marsden05}.   The errors in the region of the spectrum affected by the Moon contamination were increased by a factor of ten, although any large number would be adequate for this purpose. The profiles with the modified errors were then used as initial inputs for the Doppler imaging code. 
The pixels where the errors were increased will have less influence on the fits than the rest of the profile. Hence, the difference between the fits and the original profiles (the `difference profiles') will be dominated by the Moon contamination. This contamination (which is essentially due to sunlight) was modelled using an unspotted, non-rotating star. A spectrum of such a reference star was deconvolved (using LSD) to provide a `reference profile'. This reference profile is then fitted to each of our difference profiles.

Fitting the difference profiles involved multiplying the reference profile by a different scale factor, as the level of contamination varies depending on the weather and Moon position. To find the scale factor $\alpha$ we minimise the value of $ \Sigma_{i}(y_{i} - \alpha z_{i})^{2}$ for each profile, where $y$ is the difference profile, $\alpha$ is the scale factor and $z$ is the reference profile. 

\begin{figure}
\includegraphics[width=0.4\textwidth]{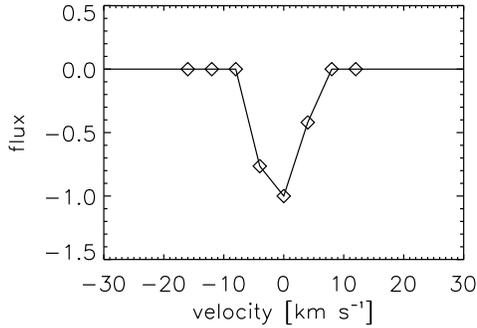}
\caption[Moon profile for TWA~17.]{The reference profile (see text) that has been multiplied by the appropriate value of $\alpha$ before being subtracted from the LSD profiles for TWA~17. }
\label{fig:inst}
\end{figure}

\begin{figure}
\includegraphics[width=0.48\textwidth]{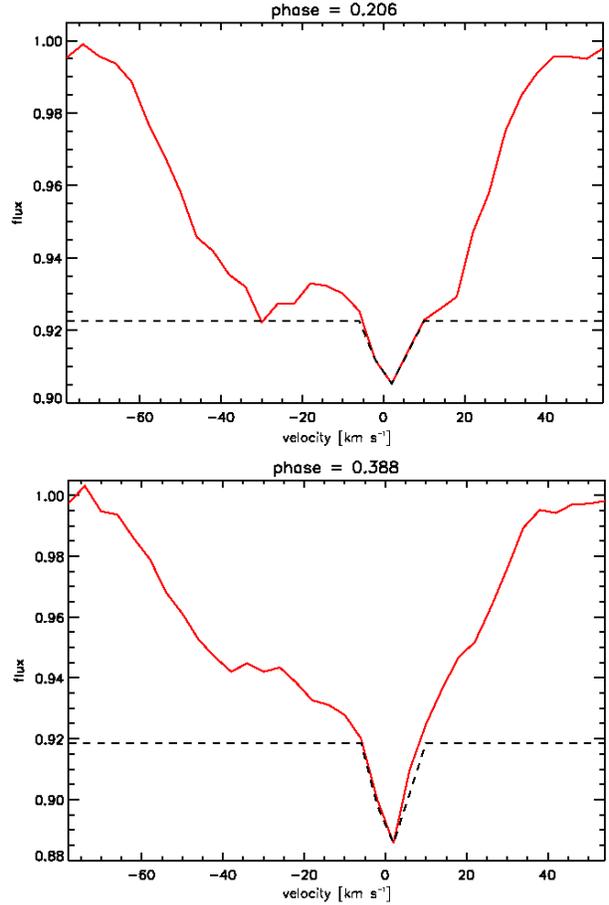}
\caption[Scaled Moon profiles plotted over two of the TWA~17 LSD profiles.]{Scaled reference profiles (dashed lines) plotted over two of the TWA~17 LSD profiles. The values of $\alpha$ are 0.017 for the top plot and 0.033 for the bottom plot.} 
\label{moonrem1}
\end{figure}

In Fig.~\ref{fig:inst} the reference profile is shown. This was created by deconvolving a template spectrum (the K3 star HD~34673) and setting the flux equal to zero at velocities where the Moon is not considered to have an effect. The flux on the vertical axis has been rescaled to give a depth of 1.0, and the continuum has been shifted to 0.0. The result of this is that the reference profile can simply be multiplied by the appropriate factor and subtracted from the LSD profiles. In Fig. \ref{moonrem1} the reference profile has been scaled in this way and overplotted on two example LSD profiles. 

\begin{figure}
\centering
\includegraphics[width=0.5\textwidth]{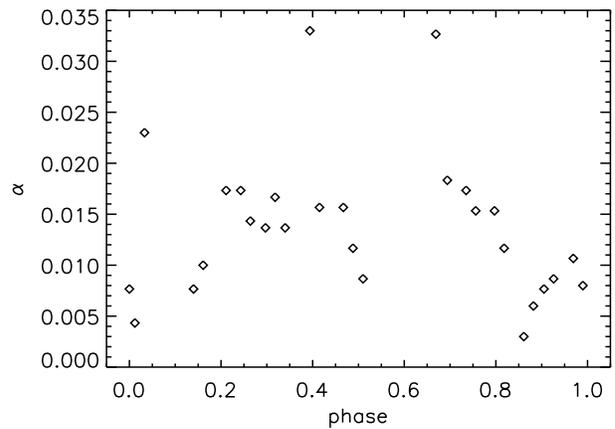}
\caption[Scale factors for the Moon removal.]{The scale factor $\alpha$, plotted against phase. Each profile $j$ has $\alpha_{j}z$ subtracted prior to reconstruction, where $z$ is the `reference profile' plotted in Fig. \ref{fig:inst}.}
\label{fig:alpha}
\end{figure}

The values of $\alpha$ found for each profile are shown in Fig. \ref{fig:alpha}. Two values of $\alpha$ are significantly higher than the others. This can be explained by the observing conditions of the individual spectra. The spectrum at phase 0.39 was taken when it was cloudy, and the spectrum at phase 0.67 had poor seeing. Finally, each profile had the difference profile, multiplied by the appropriate value of $\alpha$, subtracted. The errors used at the velocities where the contamination was removed were then set equal to the size of the difference profile.

It was noted that there is a radial velocity shift between profiles from different nights. This reached a maximum of 7.6\kms\ between the first and last nights. This shift is due to changes in the instrumental set-up from night to night. We took account of this by cross-correlating between the telluric lines in different spectra and shifting the profiles by an appropriate velocity.  

\begin{figure}
\includegraphics[width=0.5\textwidth]{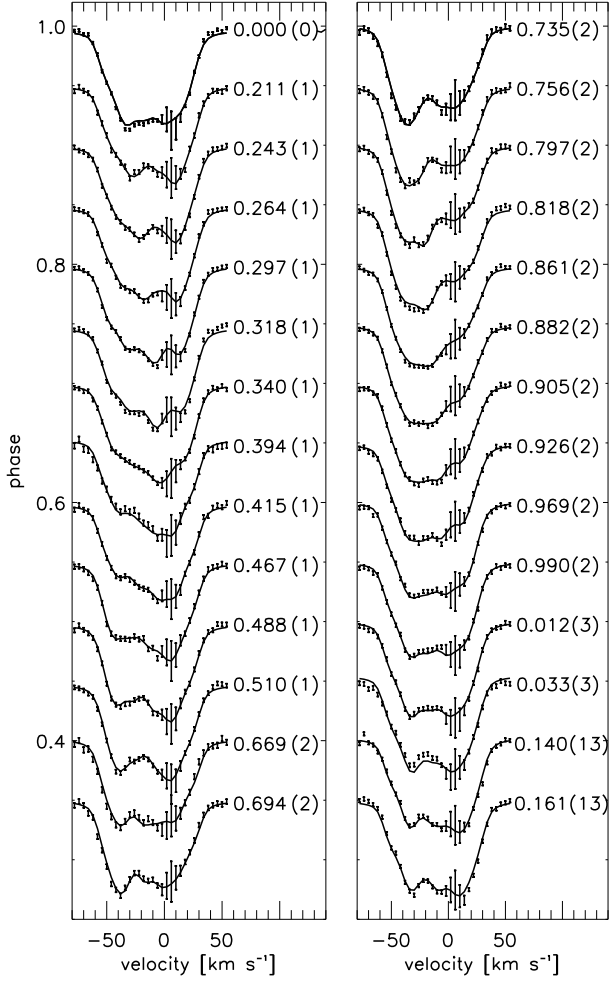}
\caption[Fits and deconvolved profiles for TWA 17] {Line profiles (1--sigma error bars) and fits (lines) for TWA~17. Errors have been increased where Moon contamination was removed. Phase and number of whole rotational periods since the first observation are shown to the right of each profile.}
\label{fig:fits17}
\end{figure}

\begin{figure}
\includegraphics[width=0.5\textwidth]{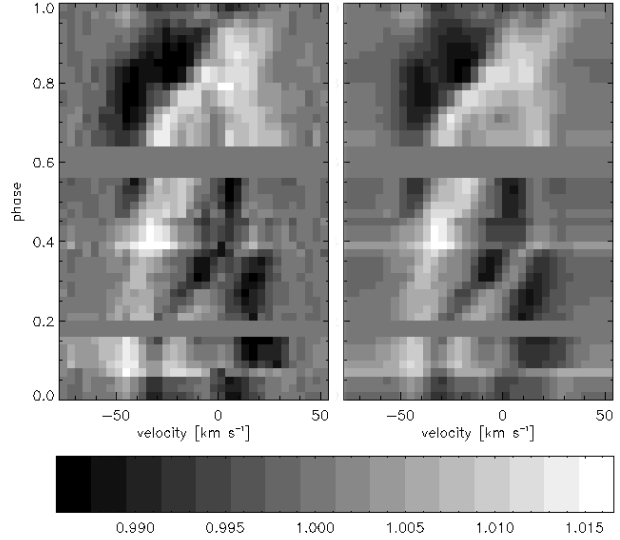}
\caption[Fits and deconvolved profiles  in greyscale for TWA 17] {Greyscale images of the LSD profiles (left) and the corresponding Doppler imaging fits (right) of TWA~17, both divided by the mean profile. Spots appear as emission (white). The greyscale is given at the bottom.}
\label{fig:grey17}
\end{figure}

\section{Doppler Imaging}\label{dopim}

\subsection{Stellar Parameters}\label{parsec}

Of the 29 LSD profiles 28 were used as input for the Doppler imaging code DoTS \citep[Doppler Tomography of Stars,][]{DoTS01}. Details of DoTS are given in Paper~I.  One spectrum was discarded as it was the last one taken on the 2006 February 12 and was therefore noisy and badly contaminated by moonlight.

First the stellar parameters, i.e., the \vsini, period, inclination angle and radial velocity were found. This was done by using DoTS to calculate \chisq\ for a range of values; the accepted value is where the minimum \chisq\ is found. Testing the code on artificial data has shown that this usually works even when the other parameters have not yet been found \citep{barnes00}.  However, the \vsini\ of the star must be determined at the same time as the equivalent width of the input spectral line. These two parameters were therefore found by simultaneously varying the \vsini\ and equivalent width of the line, and finding the combination of values that gave the lowest \chisq.

The inclination angle $i$ is the most difficult parameter to constrain as large changes in $i$ only yield small changes in \chisq.  A value of $30\dg$ gave the lowest \chisq~value with an error of $\sim10\dg$ on this value. Using the period, \vsini\ and inclination angle gives a value of $1.25^{+0.16}_{-0.13}{R}_{\odot}$ for the radius. The star can then be placed on pre-main sequence tracks in a Hertzsprung-Russell diagram of luminosity and temperature. With an effective temperature of 4400K and a luminosity of $0.53^{+0.26}_{-0.19} \rm{L}_{\odot}$ the best-fit mass from the tracks from \cite{siess00} is $1.0 \pm 0.2 \rm M_{\odot}$. All resulting and adopted parameters are listed in Table.~\ref{tab:param17}, some of which are taken from earlier work, in which case the reference is given.

%\begin{table}
%\centering
%\caption[Parameters of TWA~17]{Previously-known and updated parameters for TWA~17. References: [1]: \protect\cite{zuck04}, [2]: \protect\cite{cohen79}, [3]: \protect\cite{webb99}, [4]:\protect\cite{reid03}, [5]:\protect \cite{lawson05}, [6]: \protect \cite{dantona94}.}
%\begin{tabular*}{0.45\textwidth}{@{\extracolsep{\fill}}llll}
%\hline 
%Parameter & Previous & New Value & Reference \\
%& Value & & \\
%\hline
%Spectral type & K5 & K5 & [1] \\
%Temp. [K] & -- & $4400\pm200$ & [2] \\
%$V$ magnitude & 12.7  & -- & [4] \\
%$V~-~R$ & 1.01 & -- & [4] \\  
%Distance [pc] & 90--163pc & $120\pm40\rm pc$ & [1]\\
%\vsini~[kms$^{-1}$] & $45\pm 10$ & $47\pm 1$ & [4], this work  \\
%$v_{\rm rad}$ [\kms] & $4.6\pm6$ & $11.1\pm0.1$ & [3], this work \\
%Period [day] & $0.69\pm0.01$ &$0.685\pm0.001$& [5], this work \\
%Inclination [$^{\circ}$] & -- & $30\pm10$ & this work \\
%Radius [$\rm{R}_{\odot}$] & -- & $1.25^{+0.16}_{-0.13}$ & this work \\ 
%Mass [\mdot] & -- & $1.0\pm0.2$ & [6], this work \\
%Lum. [$\rm{L}_{\odot}$] & --  & $0.53^{+0.26}_{-0.19}$ & [4], this work \\
%Age [Myr] & 17 & $12^{+9}_{-6}$ & [5], this work \\
%\hline
%\end{tabular*}
%\label{tab:param17}
%\end{table}

\begin{table}
\centering
\caption[Parameters of TWA~17]{Previously-known and updated parameters for TWA~17. References: [1]: \protect\cite{zuck04}, [2]: \protect\cite{cohen79}, [3]: \protect\cite{reid03}, [4]: \protect\cite{mamajek05}, [5]: \protect \cite{lawson05}, [6]: \protect\cite{webb99}, [7]: \protect \cite{siess00}.}
\begin{tabular*}{0.48\textwidth}{@{\extracolsep{\fill}}llll}
\hline 
Parameter & Previous & New Value & References \\
& Value & & \\
\hline
Spectral type & K5 & K5 & [1] \\
Temp. [K] & -- & $4400\pm200$ & [2] \\
$V$ magnitude & 12.7  & -- & [3] \\
$V~-~R$ & 1.01 & -- & [3] \\  
Distance [pc] & 90--163pc & $120\pm40\rm pc$ & [1,4,5], this work \\
\vsini~[kms$^{-1}$] & $45\pm 10$ & $47\pm 1$ & [3], this work  \\
$v_{\rm rad}$ [\kms] & $4.6\pm6$ & $11.1\pm0.1$ & [6], this work \\
Period [day] & $0.69\pm0.01$ &$0.685\pm0.001$& [5], this work \\
Inclination [$^{\circ}$] & -- & $30\pm10$ & this work \\
Radius [$\rm{R}_{\odot}$] & -- & $1.25^{+0.16}_{-0.13}$ & this work \\ 
Mass [\mdot] & -- & $1.0\pm0.2$ & [7], this work \\
Lum. [$\rm{L}_{\odot}$] & --  & $0.53^{+0.26}_{-0.19}$ & [3], this work \\
Age [Myr] & 17 & $12^{+9}_{-6}$ & [5,7], this work \\
\hline
\end{tabular*}
\label{tab:param17}
\end{table}

With these parameters it was then possible to fit the data and produce Doppler images. Each pixel of the image is allowed to have a spot-filling fraction $0 < f \le 1$, and the values of the $f$ are allowed to vary as the code searches for the solution. For the photospheric temperature $T_{\rm phot}$ the pixels have $ f \rightarrow 0 $ ($f \ne 0$ to avoid singularities). Pixels at the spot temperature $T_{\rm spot}$ have $f = 1$. 
During LSD the errors for the profiles are calculated using photon noise statistics, but this process can lead to underestimation of the errors in the case of unpolarised profiles \citep{jf97b}. This leads to \chisq\ values greater than unity. 
A minimum \chisq\ of around 1.7 could be achieved but in this case we could not recover a maximum entropy image, so we relaxed the final \chisq\ requirement to 2.0.
In Fig. \ref{fig:fits17} the data are shown (with the Moon spectrum subtracted) and the fits are overplotted. The increased error bars where the Moon spectrum has been subtracted can be seen clearly and it appears that the fits are not affected by the Moon, as there are no spurious absorption features in the relevant velocity range. In Fig. \ref{fig:grey17} the data and fits are shown in greyscale, with a mean profile divided out in each case, and stacked in order of phase. The spot groups (the bright regions in the greyscale) are obvious.

\subsection{Surface Images}\label{images}

\begin{figure*}
\centering
\includegraphics[width=0.9\textwidth]{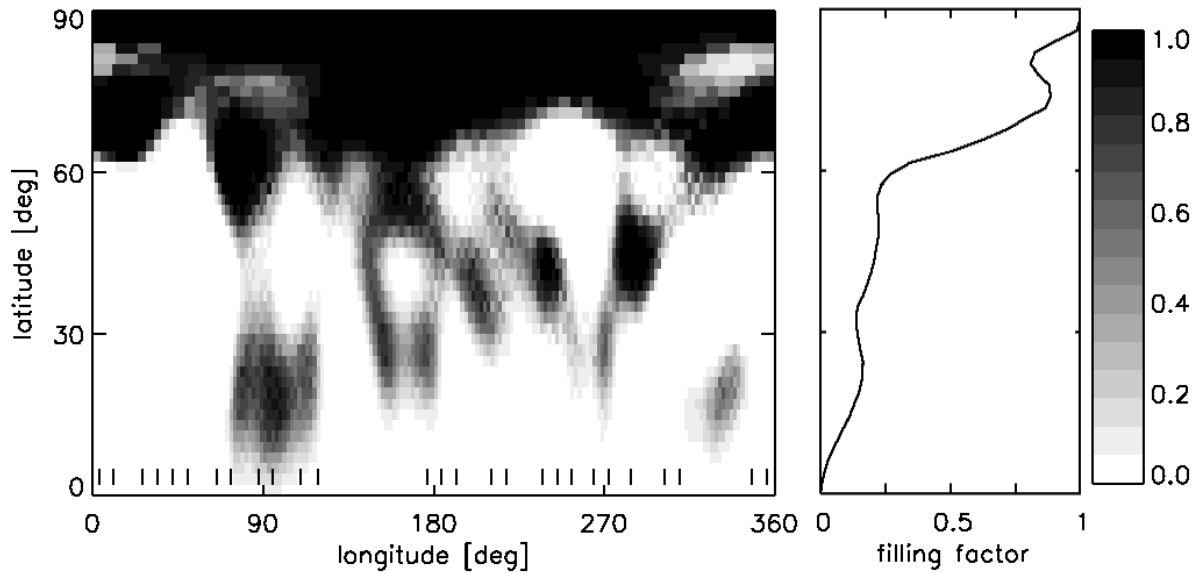}
\caption[Mercator projection of spot map of TWA~17.]{Mercator projection of the spot map produced using all the spectral data. Observation phases are indicated by the ticks at the bottom. The greyscale (given on the right) represents the filling factor. The mean spot filling factor at each latitude is also given. Phase increases from right to left, i.e., longitude of $360\dg$ is at phase 0 and $0\dg$ is phase 1.}
\label{fig:map17} 
\end{figure*}

\begin{figure}
\centering
\includegraphics[width=0.48\textwidth]{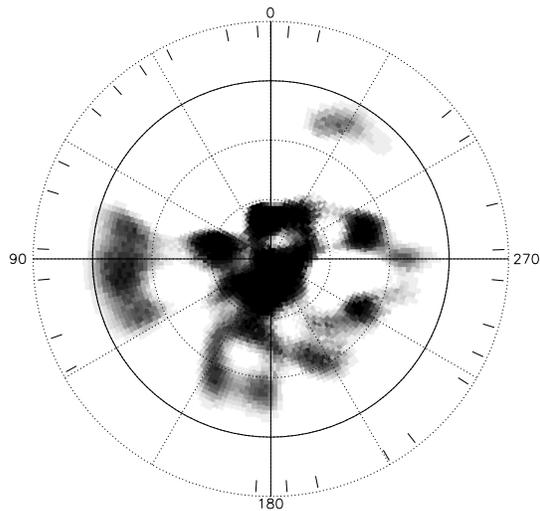}
\caption[Polar projection of the map for TWA~17.]{Polar projection of the spot map in Fig. \ref{fig:map17}. The dotted circles represent lines of latitude separated by $30 \dg$ and the solid line shows the equator. Longitudes (corresponding to the longitudes in Fig. \ref{fig:map17}) are indicated by the numbers around the outside. Observation phases are indicated by tick marks. The greyscale is the same as in Fig. \ref{fig:map17}.}  
\label{fig:polar}
\end{figure}

The map resulting from the fits to the profiles (Figs \ref{fig:fits17} and \ref{fig:grey17}) is shown in Figs. \ref{fig:map17} and \ref{fig:polar}. In common with other rapidly rotating young stars (including TWA~6) there is a polar spot and several spots at lower latitudes. The minimum and maximum temperatures indicated by the map are 3400K and 4400K respectively. This maximum temperature is determined using the spectral type (found by comparing with spectral standards). The minimum temperature is determined by decreasing the temperature until the maximum filling factor falls below 1.0.

The low SNR made it necessary to check that none of the features on the map are artefacts, caused by noise in the profiles. To do this the minimisation was repeated with random profiles removed. We wished to verify that the resulting image was similar even if we removed several profiles. However if too many profiles are removed it will leave large phase gaps. We elected to remove six randomly chosen profiles each time. 

 % latitudinal spot distributions, explanation of tickmarks 

\begin{figure*}
\centering
\includegraphics[width=0.9\textwidth]{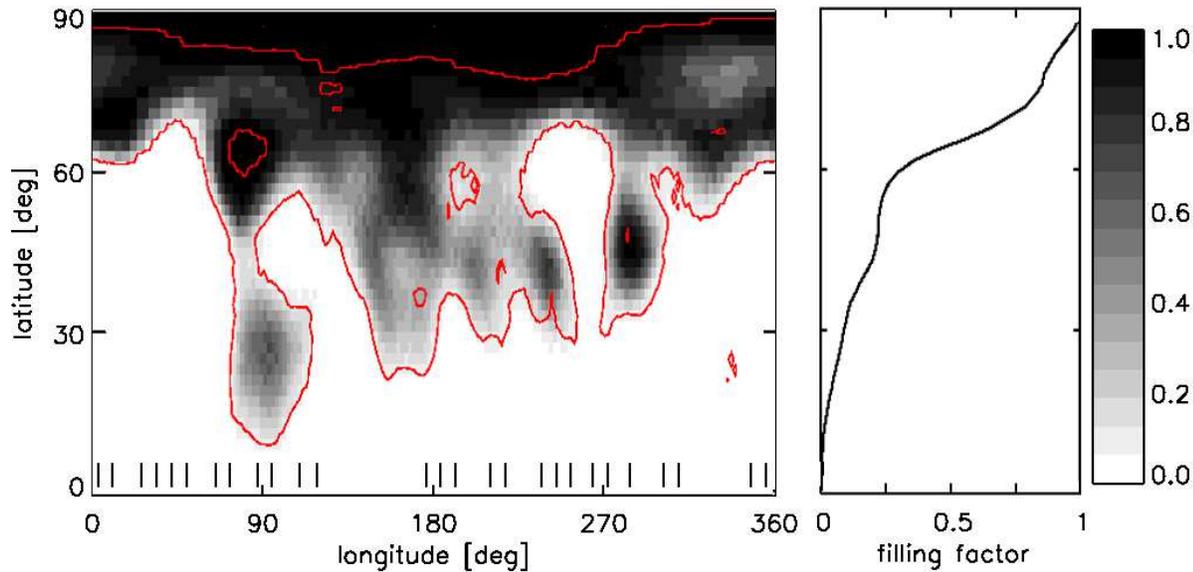}
\caption[Mean and variance image for TWA~17.]{As Fig. \ref{fig:map17} but showing the mean of all 30 spot maps (see text) in greyscale. The variance image is overplotted in red contours.   }
\label{fig:mapvar}
\end{figure*}

Once the profiles had been removed, we carried out the minimisation with 60 iterations and a target \chisq\ of 2.2. The target \chisq\ had to be slightly higher than the value of 2.0 achieved earlier to ensure the we could reach the minimum \chisq\ each time. The procedure was repeated 30 times to produce 30 separate images. The mean of the 30 images is shown in Fig. \ref{fig:mapvar} with the normalised variance image overplotted as the contours. The contours line up well with the spots. This suggests that differences between the 30 images are due to differences in the exact intensity and shape of the spots in each image, rather than their locations. The only likely artefact is the feature at $\sim 330^{\circ}$ longitude and $\sim 20^{\circ}$ latitude which appears on a small scale in a number of images and thus also in the variance image. It appears in the map produced using all data (Fig. \ref{fig:map17}) but is suppressed in the mean image. This feature is at a point where there is a gap in the phase coverage and we conclude that it is not real. 

The spot distribution is qualitatively similar to other maps of rapidly rotating PMS stars, i.e. there is a large polar spot and a lower spot coverage at all latitudes. The distribution can be compared with the flux tube models in \cite{gran2000}. The rotational velocity of TWA~17 is $\sim 50$ times solar,  and its mass is 1.0\mdot. The prediction for the spot map depends on the age we use for the star - 10 Myr for TWA or 17 Myr for LCC. In the former case TWA~17 falls into the `T' (T Tauri) classification in \cite{gran2000}, and in the latter case the `P' (pre-main sequence) classification applies. For the T case the models predict most of the emergence to be at the poles, with a smaller but non-zero spot coverage at lower latitudes.  In the P case the spots are predicted to be in a band between $30 - 60^{\circ}$. 

Our recovered map looks more like the T case, where the star is assumed to have a very small radiative core, consistent with the models of \cite{siess00}.  These models indicate that TWA~17 should have a small radiative core. However, previous Doppler images of post-T Tauri stars have also shown high latitude and polar spots, e.g. RX~J1508.6Ð4423 \citep{jf00} so as yet it is difficult to draw firm conclusions about the age of TWA~17 based on the spot distributions alone.

\subsection{Long-term evolution}\label{photo}

\begin{figure}
\includegraphics[width=0.5\textwidth]{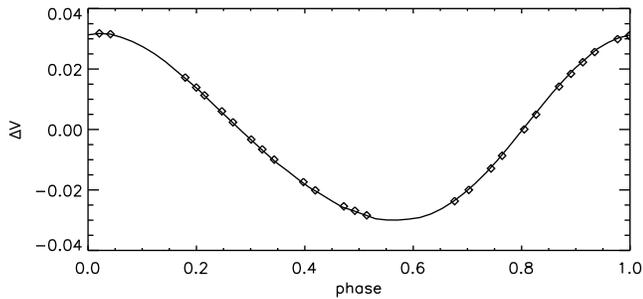}
\caption{The V-band lightcurve predicted using the intensity map in Fig. \ref{fig:map17}. The symbols show the phases where spectra were taken. }
\label{light}
\end{figure}

There is no contemporaneous photometry of TWA~17 but we can use our Doppler images to estimate the flux changes in the star as it rotates and compare with earlier photometric measurements in \cite{lawson05}, whose observations were carried out in May 2000. This is done by integrating the intensities across the visible disc of the star at each phase. The result is shown in Fig. \ref{light}. The maximum change in brightness of the star is just over 0.06 mag. In \cite{lawson05} a peak--to--peak $\Delta \rm V$ of 0.12 mag was measured. Our lower estimate is reasonable given that Doppler imaging tends to underestimate the flux changes (we discuss this in more detail in Paper 1). The lightcurve in \cite{lawson05} is more symmetrical than Fig. \ref{light},  indicating changes of the spot distribution in the intervening time.

In Paper I we found that it was not possible to recover the $\Delta \rm V$ of 0.49 mag that was measured in \cite{lawson05}, instead a value of 0.1 was calculated based on the Doppler images. We attributed this to a combination of a change in the spot coverage in the intervening years and a tendency of Doppler imaging to underestimate flux changes \citep{unruh95}. We do not find such a large discrepancy between the two TWA~17 values. This indicates that the factor of 4--5 difference in the values for TWA~6 measured in \cite{lawson05} and our calculations cannot be accounted for by Doppler imaging underestimations alone. 
\\

\section{Balmer Line Analysis}\label{balmer}

Balmer line emission is a tracer of the circumstellar environment and chromosphere of young stars such as TWA~17.  The variability in H$\rm \alpha$ and the other Balmer lines will therefore extend to velocities greater than \vsini\ if the material is in corotation with the star. 
By studying the variation of these spectral lines it should be possible to identify events, such as flaring and prominences, in the chromosphere and corona. 

\subsection{Mean Balmer line profiles}\label{sec:meanha2}

\begin{figure}
\centering
\includegraphics[width=0.48\textwidth]{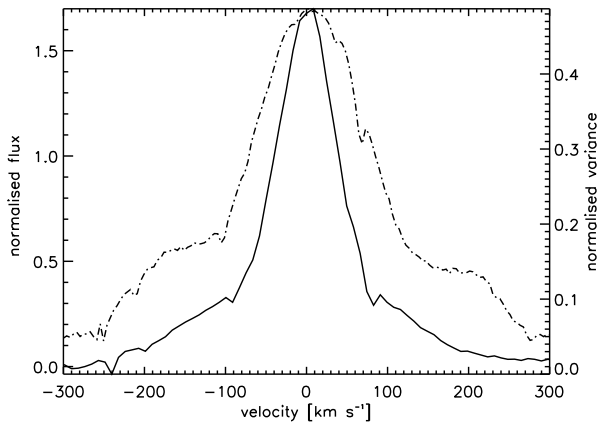}
\caption[Mean \ha\ and normalised variance for TWA~17.]{The mean H$\rm \alpha$ line profile for TWA~17 is shown as the solid line. A rotationally broadened absorption profile (calculated as described in the text) has been subtracted. The normalised variance profile for the H$\rm \alpha$ line is shown as the dashed line. The vertical axis on the left-hand side refers to the mean H$\rm \alpha$ and the right-hand axis refers to the variance profile. The small dips in both profiles are due to telluric lines. } 
\label{fig:varha17}
\end{figure}

\begin{figure}
\centering
\includegraphics[width=0.48\textwidth]{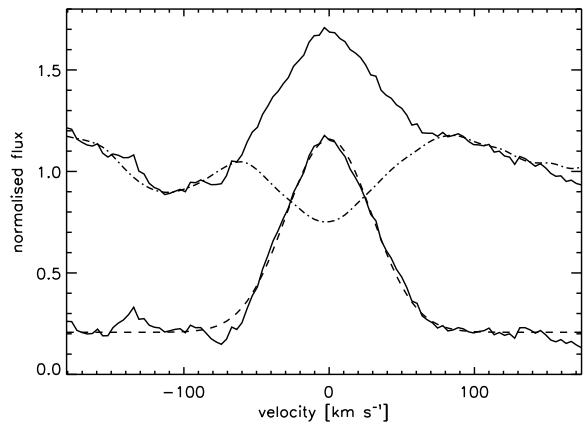}
\caption[Mean H$\rm \beta$ profile of TWA~17.]{Mean H$\rm \beta$ profile of TWA~17 before (top) and after (bottom) a rotationally broadened K5 template spectrum has been subtracted off. The broadened template spectrum is shown as the dot--dashed line. The dashed line shows the Gaussian fit to the subtracted profile.}
\label{fig:hbmean17}
\end{figure}

Figs \ref{fig:varha17} and \ref{fig:hbmean17} show the mean H$\rm \alpha$ and H$\rm \beta$ profiles respectively. The mean has been taken of the continuum normalised profiles from each spectrum (also averaging the two orders that H$\rm \alpha$ and H$\rm \beta$ each appear in). A broadened absorption spectrum has been subtracted off. The absorption spectrum was created by broadening the normalised template K7 spectrum to a \vsini\ of 47\kms\ (the \vsini\ of TWA~17 found in Section~\ref{parsec}). Also plotted in Fig.~\ref{fig:varha17} is the normalised variance profile for the \ha\ profiles, given by

%\begin{equation}
%V_{\lambda} = \Big[\frac{\sum_{i=1}^n(I_{\lambda,i}-\overline{I_{\lambda}})^{2}}{\overline{I_{\lambda}}(n-1)}\Big]^{1/2},
%\label{eq:var}
%\end{equation}

\begin{equation}
V_{\lambda} = \Big[\frac{\sum_{i=1}^n(I_{\lambda,i}-\overline{I_{\lambda}})^{2}}{(n-1)}\Big]^{1/2}\Big/\overline{I_\lambda} ,
\label{eq:var}
\end{equation}

% $$ V = \left[ \frac{\sum_{i=1}^n(I_{\lambda,i}-\overline{I_{\lambda}})^{2}}{(n-1)} \right^{1/2}] / \bar{I_\lambda} $$

\noindent where $\rm I_{\lambda,i}$ is the intensity of profile $i$ at wavelength $\rm \lambda$, $\overline{I_{\lambda}}$ is the mean of the intensities at wavelength $\lambda$ and $n$ is the number of profiles. Equation~\ref{eq:var} is based on the expression for the variance profile given in \cite{johns95}. The subtracted mean H$\rm \alpha$ profile has an equivalent width of $4.8 \rm \AA$ and a broad and narrow component. We consider the narrow component to originate in the chromosphere. The origin of the higher velocities in the broad component will be discussed later in this section. 
In Fig.~\ref{fig:hbmean17} both the subtracted and unsubtracted lines are shown because the subtraction makes a substantial difference to the shape of the line. The H$\rm \beta$ profile (with the broadened absorption spectrum subtracted off) has an equivalent width of $1.1 \rm \AA$. It can be fitted by a Gaussian profile with a FWHM of 70\kms\ (also shown in Fig. \ref{fig:hbmean17}). This is consistent with rotational broadening and a \vsini\ of 47\kms\ using the relationship between the FWHM and \vsini\ given in \cite{gray92}.  

 % overplot rotationally broadened template 

%There is no evidence of accretion. 

\begin{figure}
\centering
\includegraphics[width=0.48\textwidth]{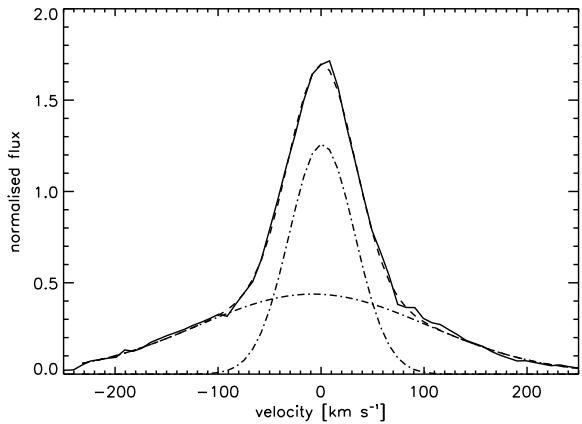}
\caption[Mean \ha\ profile of TWA~17 fitted with a broad and narrow Gaussian.]{The solid line shows the mean H$\rm \alpha$ profile.  The dashed-dotted lines are the two Gaussians that were fitted to the mean profile. They have FWHM of 75\kms\ and 263\kms. The dashed line is the sum of the Gaussians. This fit to the line has a reduced \chisq\ of 1.1.}
\label{fig:gaus17}
\end{figure}

Following the example of, e.g., \cite{petrov94} and \cite{hatzes95} we attempted to fit the mean subtracted \ha\ profiles with a broad and narrow Gaussian. To do this we used \chisq\ minimisation to find the best fitting parameters (i.e., height, width and central position) for each Gaussian. The FWHM of the narrow and broad components found using this method are 75\kms\ and 263\kms\ respectively, giving a reduced \chisq\ of 1.1. The fit is plotted in Fig.~\ref{fig:gaus17}. The largest discrepancies between the fit and the data are at the positions of the telluric lines. The width of the narrow component is in agreement with the \vsini\ of 47\kms.  There is a radial velocity shift of just under 8\kms\ between the two components, which is slightly less than one pixel in resolved velocity (after binning), and we do not judge it significant. All parameters are listed in Table~\ref{tab:gausparams}. 

\subsection{Variance profile}

The variance profile has a full-width at half-maximum (FWHM) of 178\kms\ and indicates that there is variability out to around $\pm$260\kms. There a slightly higher variability on the blue side of the emission. However, this asymmetry is not apparent in the mean H$\rm \alpha$ profile. The narrow component of the variance profile is wider than the narrow component of the mean profile. As the wings may vary more than the line centre the full-width at zero intensity (FWZI) may be the more relevant parameter. We estimate that the FWZI is 260-280\kms\ although it is difficult to determine due to the broad `pedestal' component.
The normalised variance profile for the H$\rm \beta$ profile has essentially the same shape as the mean profile but is quite noisy and is not shown. %(var_hb.ps) 

\begin{table}
\centering
\caption[Parameters of gaussians used to fit \ha\ lines.]{Parameters of Gaussians used to fit \ha\ lines, for the narrow and broad components.}
\begin{tabular*}{0.48\textwidth}{@{\extracolsep{\fill}}llll}
\hline
Component & Amplitude & Centroid & Width \\
 & & [\kms] &  [\kms] \\
\hline
Narrow & 1.3 & 1.0 & 75  \\
Broad & 1.0 & -7.8 & 263  \\
\hline
\end{tabular*}
\label{tab:gausparams}
\end{table}

In previous work, e.g., \cite{fern04}, the broad `pedestal' part of the emission has been explained by microflaring, i.e., small flares that cannot be resolved individually. The motions of the material within these flares can explain the emission seen in the wings of the line profiles. Microflaring may therefore be a contributor to the variability in the H$\rm \alpha$ line.  

\subsection{Variability in the Balmer lines} 

\begin{figure}
\includegraphics[width=0.5\textwidth]{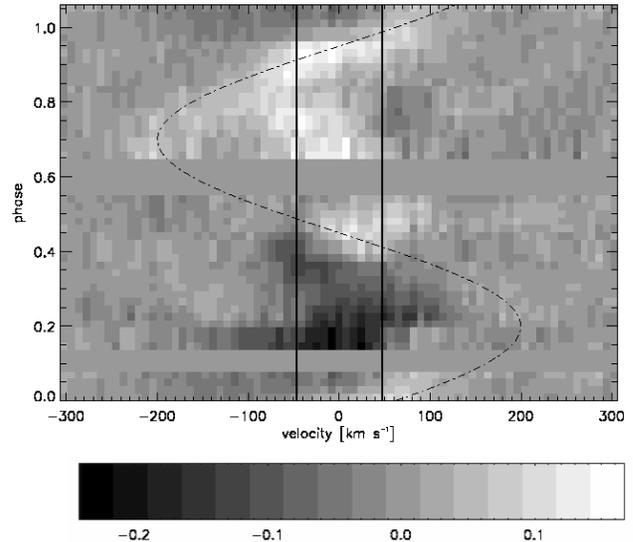}
\caption[H$\rm \alpha$ profiles of TWA~17 shown in greyscale.]{H$\rm \alpha$ profiles of TWA~17 stacked, with a mean profile subtracted, shown in greyscale. The scale bar is shown at the bottom. The vertical lines indicate $\pm$\vsini. The dash-dot curve is a sinusoid indicating the position of the feature identified as a slingshot prominence (see text). The vertical axis shows phases from 0 -- 1.05, in order to show the continuation of the bright feature.  }
\label{fig:havar17}
\end{figure}

In Fig. \ref{fig:havar17} the greyscale image of the H$\rm \alpha$ variability is shown. This was produced by stacking all the \ha\ emission lines in order of phase and subtracting the mean \ha\ profile (Fig. \ref{fig:varha17}). The greyscale at each pixel therefore represents the difference in flux between that pixel and the mean, hence a greyscale value of zero indicates the emission at that pixel is equal to the mean. The velocity scale in Fig. \ref{fig:havar17} has been shifted to take account of the radial velocity. The asymmetry that was evident in the normalised variance profile (see Fig.~\ref{fig:varha17}) is also visible in Fig.~\ref{fig:havar17} as the left (blue) side of the image has more features (i.e bright or dark regions) than the right (red) side. The equivalent \hb\ image was very noisy (as it is at the blue end of the spectrum where there are fewer counts) and therefore did not show any clear features. Consequently the \hb\ image is not shown here. 

On TWA~6 we observed signatures of active regions that appeared to be cospatial with the spots found in the Doppler imaging. However, there is no evidence of such an alignment between the spots and active \ha\ regions on TWA~17.

%the greyscale has the range -0.2-0.1 This means that there is more absorption than emission. However, as a mean proÞle has been subtracted from each proÞle it may be that within ±v sin i these proÞles have the base level of emission. If that is the case then higher emission at other phases causes these pixels to appear dark once the mean has been subtracted. 

Variability in \ha\ can be due to scattering of \ha\ photons from the stellar surface by circumstellar hydrogen clouds. If these clouds are in front of the stellar disc, photons will be scattered out of the line-of-sight. This leads to reduced emission, which will appear in images such as Fig. \ref{fig:havar17} as dark regions. On the other hand if the clouds lie off the stellar disc photons will be scattered into the line-of-sight leading to increased emission. 
Variability at velocities above \vsini\ can either be caused by high-lying material in co-rotation with the star (e.g., slingshot prominences, see Section \ref{intro}), or fast-moving material closer to the surface, e.g. related to flares.

If a slingshot prominence is present and observed at all phases it should be possible to fit a sinusoid to the signatures. It may pass in front of the stellar disc, depending on its height and inclination, in which case it will most likely cause a decrease in emission. This will appear as a dark diagonal band going from blue to red as the material is in corotation with the star. 
Scattering of  H$\rm \alpha$ photons can also be caused by flare material, and infalling gas, which may also be accompanied by velocity changes that can indicate the direction of motion of the material. 

Studying Fig.~\ref{fig:havar17} we see two features that may be explained by these phenomena. The first of these is a bright region between $\pm 80$\kms\  and phases $\sim 0.4 - 0.8$ going from red to blue, which may indicate that it is material on the far side of the star that does not go out of sight due to the inclination of the system. The second feature is a prominent brightening between 0.65 and 1.075 that extends between -220\kms\ and approximately +120\kms. It is possible that this is a slingshot prominence of the type discussed above, although this is difficult to confirm as we do not observe it for a whole rotation period. In Fig. \ref{fig:havar17} this brightening is overlaid with a sinusoid that represents the expected path of a prominence of this size. As the observations that gave these spectra were all taken in one night (2006 February 13) we have no handle on its lifetime. It does not appear in absorption between $\pm$\vsini, which may indicate that it is indeed a prominence that lies at a distance of more $3 R_{*}$ above the surface.

In summary, although microflaring may be a contributor to the variance, the large scale features that have high velocities in Fig.~\ref{fig:havar17} are important contributors to the overall variability in TWA~17. 

\section{Distance to TWA~17 and the TW~Hydrae Association} \label{comp}

In this section we compare our findings for TWA~6 and 17 and consider the question raised in Section \ref{intro} of whether they have similar distances and ages and hence can be considered members of the same association. 

Using the parameters found in Section \ref{parsec} the radius of TWA~17 was estimated as $1.25^{+0.16}_{-0.13}\rm{R}_{\odot}$ compared with $1.05^{+0.16}_{-0.15}\rm{R}_{\odot}$ for TWA~6. If TWA~6 and TWA~17 were to belong to the same association and were thus at a similar distance, we would not expect TWA~17 to have a higher effective temperature and be fainter than TWA~6. 

However, this discrepancy can be resolved if we consider that TWA~17 may be more distant than TWA~6. For the parameters determined in this paper (specifically the radius) to be consistent with the spectral type and luminosity measured previously a distance of $190 \pm 130 \rm pc$ would be required, where the error is dominated by the error in the radius that has been propagated. Given that TW Hya has a Hipparcos distance of 56 pc this might suggest that TWA~17 lies in the LCC subgroup, although the magnitude of the errors makes it difficult to draw firm conclusions on this point,

Previous estimates for TWA~17 have suggested distances between 100 and 150 pc.  For instance, in \cite{lawson05} two separate populations have been identified within the stars previously identified as members of the TWA, those numbered 1 -- 14 are the `true' TW Hydrae association and the others are members of the more distant Lower Centaurus--Crux (LCC) subgroup of the Ophiuchus--Scorpius--Centaurus association, at an average distance of 90 pc. An even more distant estimate is given in \cite{mamajek05} where the distance to TWA~17 is $163\pm46$ pc. In addition \cite{zuck04} give a luminosity distance of 133 pc for TWA~17 and conclude based on this  that it is a member of the LCC. The results from this paper are consistent with this interpretation. 

\begin{figure}
\includegraphics[width=0.47\textwidth]{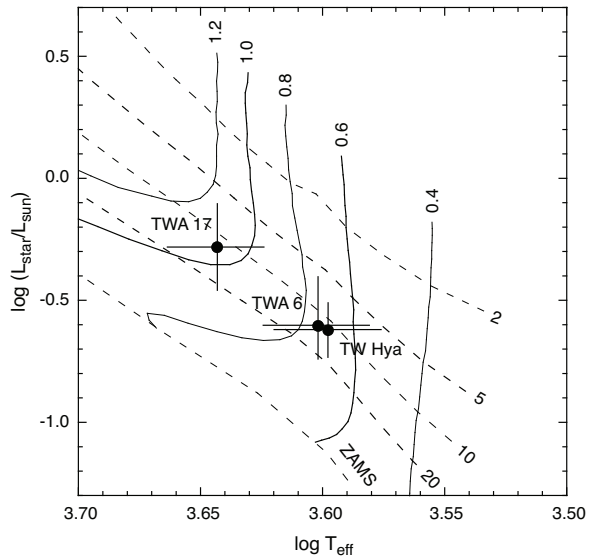}
\caption{Luminosity against effective temperature for TWA~6, TWA~17 and the brightest star in the association, TW Hya. The solid lines are model evolutionary tracks in units of solar masses. The dashed lines show isochrones annotated in units of millions of years. The tracks are from \protect\cite{siess00}. }
\label{hrd}
\end{figure}

Using the luminosity and temperature values which we have adopted for these two stars we have placed them in a Hertzsprung-Russell (HR) Diagram in Fig. \ref{hrd}, along with the brightest star in the association: TW Hya. The parameters for TW Hya come from previous work  using a spectral type of K8 \citep{torres00}, and a luminosity of $0.24\rm L_{\odot}$ \citep{yang05}. The tracks are from \cite{siess00}. All three stars fall closely below the 10 Myr isochrone. 

We have looked at several other age indicators for the two stars. Both have Li~$6707 \rm \AA$ absorption. TWA~6 has a Li $6707~\rm \AA$ equivalent width of $0.6\rm\AA$ and TWA~17 has an equivalent width of $0.5\rm\AA$, suggesting that TWA~6 may be slightly younger. Curiously, we measure a Li EW of $0.35\rm\AA$ for TW Hya from our reliable spectra which would imply that TW Hya is older. However, the lower equivalent width in TW Hya might be a consequence of photospheric veiling due to disk accretion. The difference in effective temperature can also lead to a spread in the lithium EW of stars of identical ages \citep{soderblom93}.

TWA~6 and TWA~17 both have H$\rm\alpha$ equivalent widths of around $\sim 5 \rm\AA$ which classifies them as WTTS according to \cite{barrado03}. The Doppler images of both stars are similar, both have a polar spot and spots at all latitudes to the equator. As stated in Section \ref{images} this corresponds most closely with the T Tauri models in \cite{gran04} (rather than post-T Tauri or ZAMS), but it is not possible to draw firm conclusions on the star's age based on this.

The LCC stars are believed to be older than the TWA. Although these results presented here appear to show that the stars are the same age we cannot rule out the possibility that TWA~17 is older than TWA~6 as the error bars in Fig. \ref{hrd} encompass a wide range of ages.

\section{Differential Rotation}

Due to the observing conditions in the second half of the observing run the observations concentrated on TWA~6, making it difficult to measure differential rotation on TWA~17. We can nevertheless use the Doppler imaging code with the differential rotation parameter $\Delta\Omega$ as an input, and use \chisq\ minimisation to attempt to place an upper limit on its value \citep[see e.g.,][]{cc02}. 

When reduced \chisq\ is plotted as a function of angular velocity $\Omega$ and $\Delta\Omega$, the resulting surface is noisy. The minimum contours form an extended ridge that crosses the $\Omega - \Delta\Omega$ surface diagonally and encompasses a large range of differential rotation values. It is thus difficult to determine the confidence limits for the differential rotation of TWA~17. This is not surprising as we have only two profiles in the latter half of the run, and the phases do not correspond to any of the phases in the earlier part of the observing run. Furthermore, the relatively low inclination of the star ($\sim 30^{\circ}$) reduces the resolution at lower latitudes allowing a wider range of differential rotation values to fit the data. The period quoted in Table~\ref{tab:param17} assumes zero differential rotation. 

As stated in Section \ref{comp}, TWA~17 should be an interesting candidate for a more accurate differential rotation measurement. This is due to its position on the HR diagram (Fig.~\ref{hrd}) at the turn-off from the Hayashi to the Henyey tracks, according to the \cite{siess00} models.

\section{Conclusions}

We have presented Doppler images of TWA~17. A comparison with TWA~6 (Paper~I) shows broadly similar features, i.e. a polar spot and further spots extending out of it down to the equator. The latitudinal distribution of the spots is similar to that predicted by the models in \cite{gran04} for a rapidly rotating T Tauri star.     

Despite the presence of substantial absorption features in the spectrum due to contamination by Moonlight, and relatively low SNR we are confident that the map obtained is robust, as the minimisation was repeated a number of times with random phases removed, giving similar maps each time.  

We also studied the chromospheric emission. The mean H$\rm \beta$ emission is consistent with rotational broadening, once a template absorption spectrum has been subtracted off. H$\rm \alpha$ has a broad component in addition to the narrow component. The presence of a broad component may indicate micro-flaring, although this cannot explain all of the variability in the H$\rm \alpha$ profiles. 
The H$\rm \alpha$ line shows variability at velocities greater than \vsini, indicating that there is fast--moving material in the chromosphere, possibly flares or prominences. There is at least one feature in the emission that may be caused by a prominence. It lies at a distance of over $4 R_{*}$ from the axis of rotation. However it is only apparent in one night's observations, making it difficult to draw conclusions about the lifetime. In Paper I we found evidence for chromospheric active regions that are co--spatial with the photospheric spots on TWA~6, but this work does not find any evidence for such regions on TWA~17.

We obtain a distance of $190\pm130$ pc for TWA~17, which suggests that it does not lie in the TW Hydrae association but in a more distant subgroup. However, the age estimators indicate that TWA~17 and the bona fide association member TWA~6 are approximately of the same age. 

In the Balmer line analysis, there are signatures of high-lying prominences with lifetimes of several rotational periods. In Paper~I at least one prominence was visible on TWA~6. The increased emission at velocities outside $\pm$\vsini\ suggest that similar features exist in TWA~17. The fact that we see evidence of such features in both stars would justify a dedicated survey of a young stellar cluster to gain a better picture of the frequency and nature of slingshot prominences in T Tauri stars. 

\section*{Acknowledgments}

Thanks to the referee Uwe Wolter whose comments greatly improved the clarity of the manuscript. Thanks to N. Dunstone for useful discussions; the staff at the AAT for their help during the observation run and the Vienna Atomic Line Database for providing the linelists. Data reduction was carried out using the {\sc STARLINK} package. MBS acknowledges the support of an STFC studentship. WAL acknowledges financial support from UNSW@ADFA Faculty Research Grants. 

\bibliographystyle{mn2e}
\bibliography{references,bibliography}

\label{lastpage}
\end{document}